\documentclass[12pt]{article}
\usepackage{amsmath}
\usepackage{graphicx}
\usepackage{amssymb}
\usepackage{float}
\usepackage{bm}
\usepackage{color}
\usepackage{authblk}

\begin{document}
\title{Cold Crystal Reflector Filter Concept}
\author{G. Muhrer}
\affil{Los Alamos National Laboratory, Los Alamos, 87545, NM, USA}

\begin{abstract}
In this paper the theoretical concept of a cold crystal reflector filter will be presented. The aim of this concept is to balance the shortcoming of the traditional cold polycrystalline reflector filter, which lies in the significant reduction of the neutron flux right above (in energy space) or right below (wavelength space) the first Bragg edge. 
\end{abstract}

\maketitle

%
%
\section{Introduction}
The cold beryllium reflector filter concept to enhance the cold flux (energies below 5~$meV$) of a neutron moderator was first published by J.~M.~Carpenter et al~\cite{Carpenter1981}. This concept was also experimentally investigated by Y.~Kiyanagi et al~\cite{Kiyanagi1995} and by E.~J.~Pitcher et al~\cite{Pitcher2003}. However these measurements did not reproduce the predicted values from particle transport codes like MCNPX~\cite{MCNPX}. The reasons for these disagreements, at least in the case of~\cite{Pitcher2003}, were then revealed by Muhrer et al~\cite{Muhrer2007}. Based on this understanding the concept of the cold beryllium reflector was implemented into the Mark-3 version~\cite{Muhrer2005},~\cite{MockoMark3-pre} of the production target of the Manuel Lujan Jr. Neutron scattering Center~\cite{Lisowski2006}. The performance of the cold beryllium reflector filter concept was confirmed by Mocko et al~\cite{MockoMark3-post}. The characteristic of the cold beryllium reflector-filter concept is that flux increase below 5~$meV$ goes hand in hand with a significant flux decrease immediately above 5~$meV$. While for the vast majority of experiments performed on small angle scattering instruments and reflectometers, neutrons above 5~$meV$ are of very little interest, for some experiments, like those involving liquids on reflectometers~\cite{SPEAR},
neutrons with a wavelength below 4~\AA~are of essence. This paper will discuss a concept that will preserve the cold flux, which was gained by the reflector filter concept, while regaining the flux below 4~\AA~which was lost because of the reflector filter concept.

%
%
\section{\label{sec:2}Present status}

\begin{figure}
    \includegraphics[width=12cm]{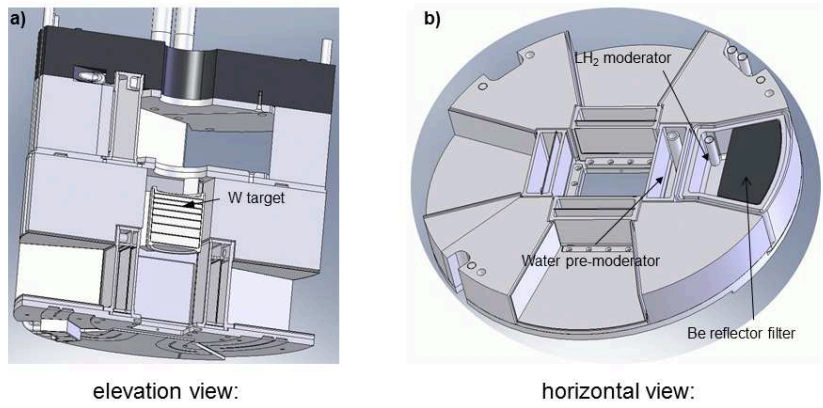}
    \caption{Vertical~(a) and horizontal~(b) cross-sections of the Mark-3 version of the spallation source of the Manual Lujan Neutron Scatting Center~\cite{Urban2012}.}
   \label{fig:Mark3}
\end{figure}

\begin{figure}
   \includegraphics[width=12cm]{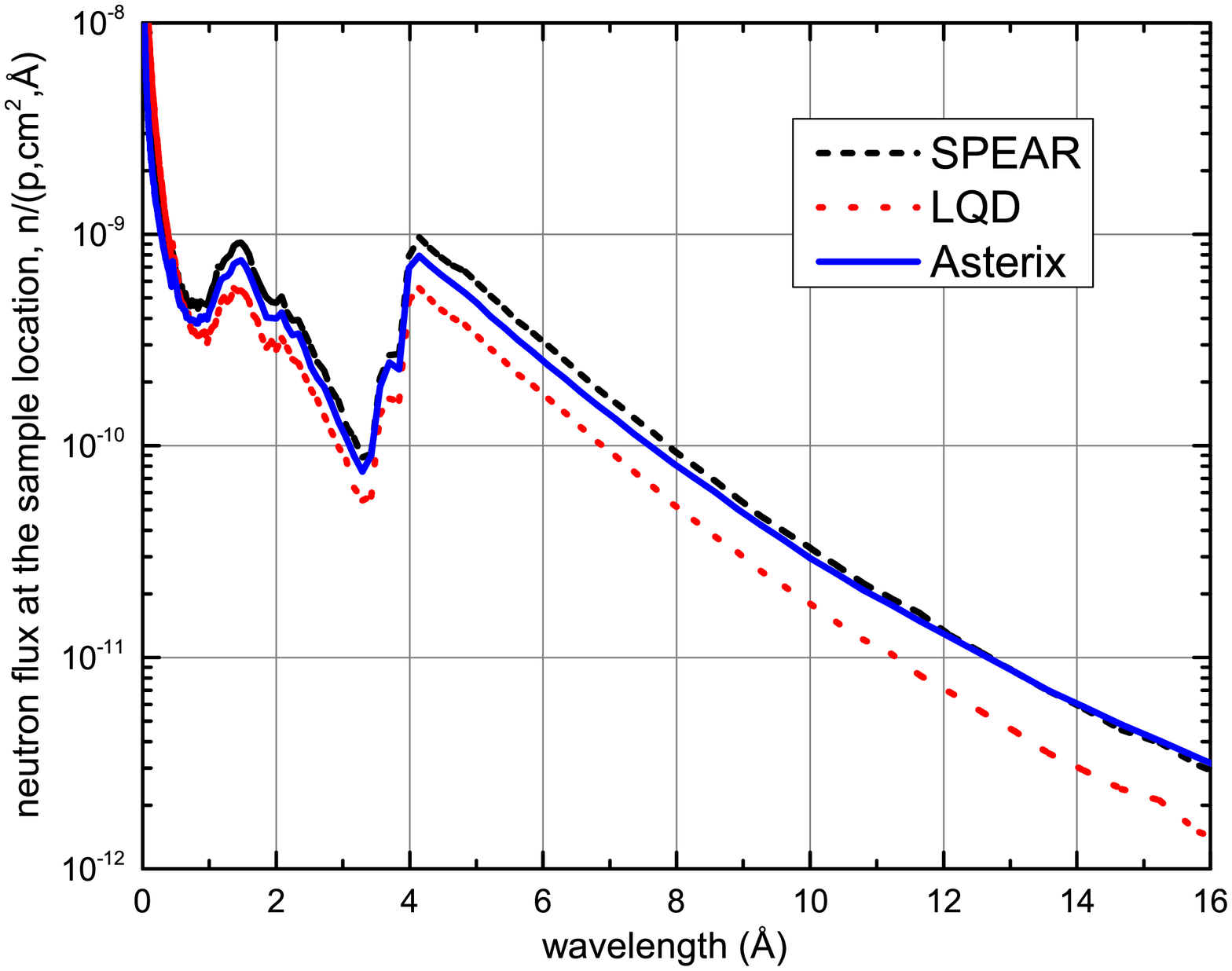}
   \caption{Calculated spectra at the sample locations as observed by the instruments viewing the cold beryllium reflector filter moderator.}
   \label{fig:Mark3-Spectrum}
\end{figure}

The cold beryllium reflector, as implemented in the Mark-3 version of the production target of the Manuel Lujan Jr. Neutron Scattering Center, is shown in figure~\ref{fig:Mark3}.
This moderator serves three neutron scattering instruments, the small angle scattering instrument LQD~\cite{LQD} and the two reflectometers SPEAR~\cite{SPEAR} and Asterix~\cite{Asterix}. The neutron spectra these instruments observe at their sample locations can be seen in figure~\ref{fig:Mark3-Spectrum}.
As it can be seen from this figure the flux decreases significantly at all flight paths above 3.96~\AA (right below 5.2~$meV$), which corresponds to the lowest Bragg edge of beryllium in energy space. This makes it difficult to measure free liquids on SPEAR. In accordance with~\cite{SPEAR},~\cite{Jarek} the following quote can be stated:
Under the boundary conditions of the SPEAR instrument geometry the scale of the momentum transfer vector $\vec{Q}$ can be written as:
\begin{equation}
   Q=4 \pi \frac{sin(\varTheta)}{\lambda},
\end{equation} 
with $\Theta$ being the scattering angle and $\lambda$ being the wavelength of the incoming neutron.
The inclination of the SPEAR beam to the horizon is about 1$^{\circ}$. In addition in order to avoid artifacts from the discontinuity caused by the beryllium Bragg edge, only data with a wavelength longer than 4.5~\AA~are used, which corresponds to $Q_{max}=0.05$ \AA$^{-1}$. If the wavelength band were extended to 1.5~\AA, the maximum of the scale of the momentum transfer vector would be $\bar{Q}_{max}=0.15$ \AA$^{-1}$. In real space the $Q_{max}$ value corresponds to a length scale of 63~\AA. This means that in the current configuration only length scales longer than 63~\AA~can be probed in free liquids. In the case of the $\bar{Q}_{max}$ this length scale would be reduced to 21~\AA, which is more in agreement with the length scales usually expected in free liquids.
In order to re-enable this type of science on SPEAR, it is desirable to regain at least part of the flux that was lost below 4~\AA~due to the implementation of the reflector filter concept.

%
%
\section{Cold Crystal Reflector Filter Concept}

At the core of the cold crystal reflector filter concept is the idea that at least part of the currently deployed polycrystalline beryllium reflector filter will be replaced by a single crystal. 
The difference between these two concepts is that in case of the perfect polycrystalline material the coherent elastic scattering (Bragg scattering) properties are independent of the vector of the incoming neutron. 
At the same time, a single crystal can be oriented in a way that 
none of the Bragg conditions are fullfilled in transmission direction
and the transmission cross section is only composed of the inelastic and the absorption cross section. In order to minimize the inelastic cross section the Debye Waller factor has to be minimized, which can be achieved by reducing the temperature to cryogenic temperatures. 
While the absorption cross section needs to be as small as possible too, temperature does not impact its value. In addition the material considered needs to be easily grown into a single crystal. Based on these boundary conditions, one very obvious choice is graphite.

\subsection{\label{sec:3} Graphite Scattering Kernel}

\begin{figure}
   \includegraphics[width=12cm]{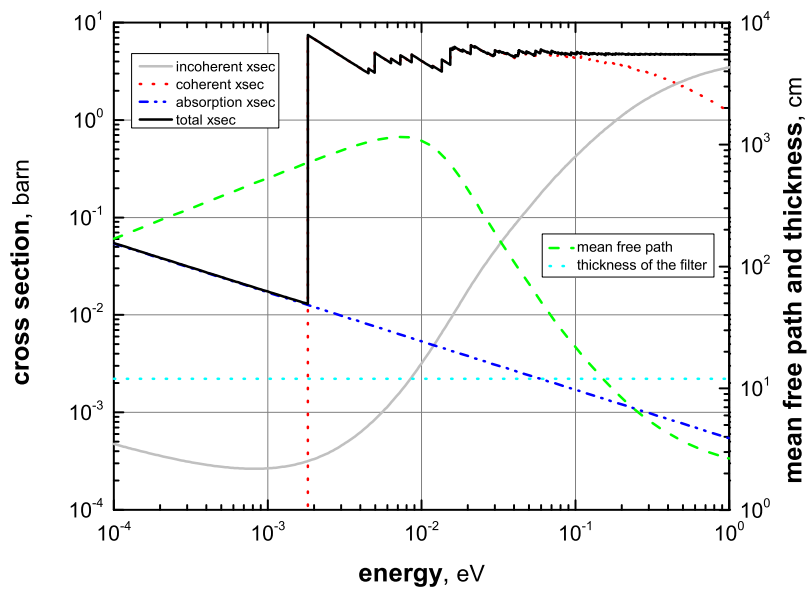}
   \caption{Theoretical cross section of polycrystalline graphite at 20~K, mean free path through a graphite crystal oriented in off Bragg conditions and thickness of the crystal reflector filter.}
   \label{fig:graphite20K}
\end{figure}

Polycrystalline graphite is very commonly used in nuclear reactor applications. It is one of the materials for which cross-section files containing the scattering laws for Monte Carlo particle transport calculations have been developed. These so called scattering kernels are usually computed using the nuclear physics code NJOY~\cite{njoy1994}. Ref. \cite{njoy1994} also contains an input deck for polycrystalline graphite at 300~K. This input deck was modified to produce a scattering kernel for polycrystalline graphite at 20~K (figure~\ref{fig:graphite20K}). 
It can be seen in this figure that in the thermal region the coherent elastic scattering is the dominant process. 
Therefore, if a graphite crystal at this temperature is oriented in off Bragg conditions, this will increase the mean free path in transmission direction significantly. Figure~\ref{fig:graphite20K} shows 
that for a 12~cm thick graphite crystal even at 200~meV the mean free path is at about 10~cm.
The main numerical problem lies in the fact that commonly used Monte Carlo particle transport codes like MCNPX~\cite{MCNPX} do not support particle transport within a single crystal. Since in this particular case one is primarily only interested in the transmission through the crystal in an off Bragg condition, the assumption is made that if one eliminates the coherent elastic scattering from the scattering kernel, it will be a good first order approximation for the single crystal. Technically, this is accomplished by reducing the values of the coherent elastic scattering cross section in the LEAPR output~\cite{njoy1994} by 10 orders of magnitude. This effectively eliminates the coherent elastic scattering while keeping the structure of the cross section file the same, which eliminates the need for re-writing the source code of MCNPX. The applicability of this assumption was confirmed by F.~X.~Gallmeier~\cite{Gallmeier2012}, who developed an first alpha test version of MCNPX which can transport particles within a single crystal. However this version is not yet available to beta testers or general users of the code.

\subsection{\label{sec:Mark-4} Mark-4 preliminary design}

\begin{figure}
  \begin{center}
   \includegraphics[width=12cm]{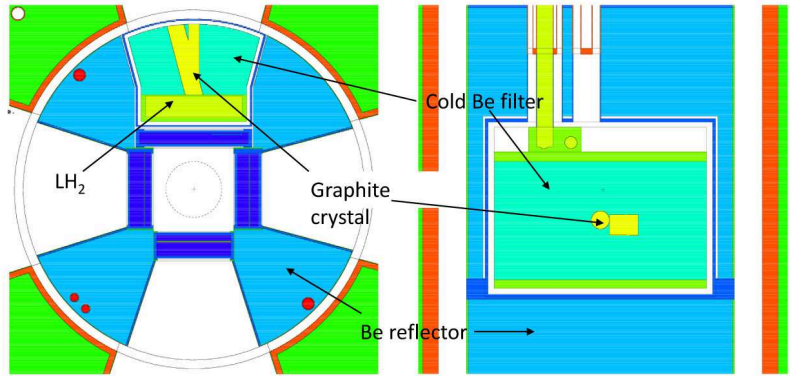}
  \end{center}
   \caption{Preliminary Mark-4 design. Two graphite crystals are added to the current Mark-3~\cite{MockoMark3-post} design.}
   \label{fig:Mark-4}
\end{figure}

In cooperation with the instrument scientists of Asterix, LQD and SPEAR, design metrics for the Mark-4 version of the production target at the Manuel Lujan Jr. Neutron Scattering Center have been developed. The simplified versions of these metrics state that SPEAR and LQD want their flux between 1.5~\AA~and 4~\AA~increased while maintaining their flux above 4~\AA. On the other hand Asterix is mainly interested in maintaining the flux above 4~\AA~and considers neutrons with a shorter wavelength primarily as background. It was therefore the goal of the initial investigation to keep the flux for Asterix the same as with Mark-3. In order to compromise between these metrics the cross-section of the graphite crystal facing Asterix was minimized. Figure~\ref{fig:Mark-4} shows the layout of a preliminary Mark-4 design.

\begin{figure}
   \includegraphics[width=12cm]{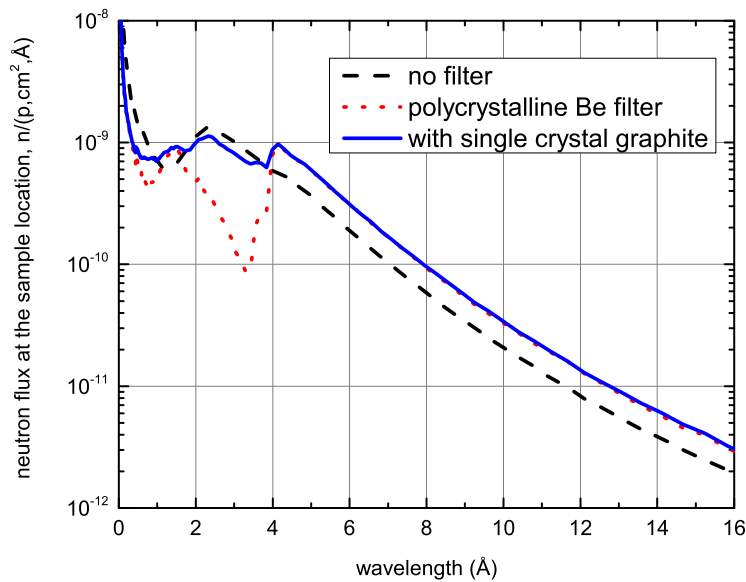}
   \caption{Calculated spectrum at the SPEAR instrument sample location. Without Be filter (dashed), with Be filter (doted) and with graphite crystals (solid)}
   \label{fig:SPEAR}
\end{figure}

\begin{figure}
   \includegraphics[width=12cm]{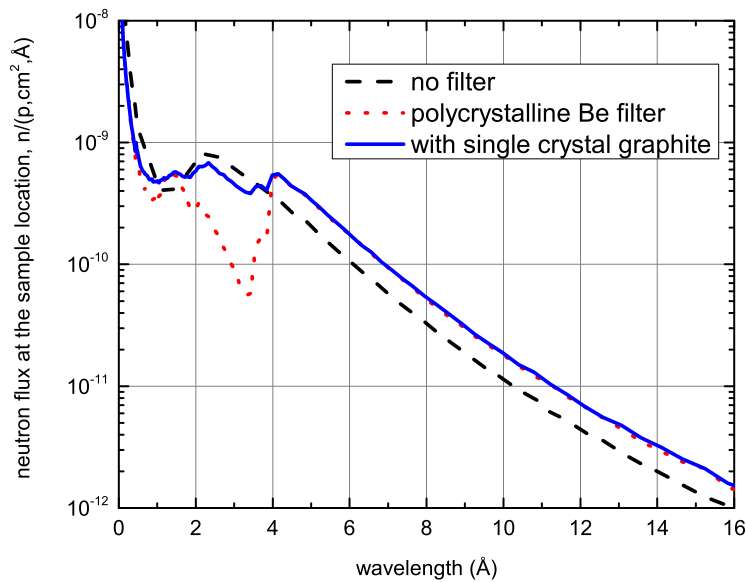}
   \caption{Calculated spectrum at the LQD instrument sample location. Without Be filter (dashed), with Be filter (doted) and with graphite crystals (solid)}
   \label{fig:LQD}
\end{figure}

\begin{figure}
   \includegraphics[width=12cm]{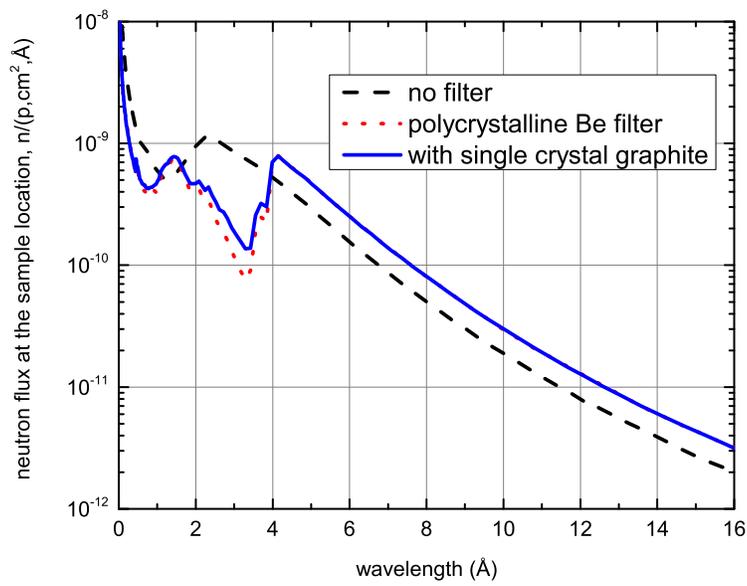}
   \caption{Calculated spectrum at the Asterix instrument sample location. Without Be filter (dashed), with Be filter (doted) and with graphite crystals (solid)}
   \label{fig:Asterix}
\end{figure}

The graphite crystals are placed in a way that there is a direct line of sight between the liquid hydrogen (LH$_2$) and the instruments SPEAR and LQD through each graphite crystal, respectively. The size of the graphite pieces were chosen to cover the core of the areas viewed by these instruments. In direction of the LQD instrument the graphite is cylindrical with a radius of 1~cm. In direction of the SPEAR instrument the cross section of the graphite is rectangular with a height of 2.5~cm and a width of 3.0~cm. The results of the simulation can be seen in figure~\ref{fig:SPEAR},~\ref{fig:LQD} and~\ref{fig:Asterix}.
Figure~\ref{fig:SPEAR} shows that by embedding the graphite crystal in the beryllium, the SPEAR instrument would be able to regain almost all the flux between 1.5~\AA~and 4.0~\AA~that was originally lost due to the introduction of the cold beryllium reflector filter. Nevertheless, there is still a significant discontinuity at 4~\AA~which will cause difficulties in the data analysis. 
As it can be seen in figure~\ref{fig:LQD}, the LQD instrument, similar to the SPEAR instrument, will regain most of its flux between 1.5~\AA~and 4.0~\AA, but still showing a significant discontinuity at 4~\AA.
The high energy neutron flux, which is the main source for the background at these instruments will remain at the same level as with the solid beryllium reflector filter. 
Finally, figure~\ref{fig:Asterix} shows that in the case of the Asterix instrument the proposed changes to the design will make very little difference to the observed spectrum at the Asterix sample location.

\subsection{\label{sec:Radius} Dimension of the graphite crystal}

\begin{figure}
   \includegraphics[width=12cm]{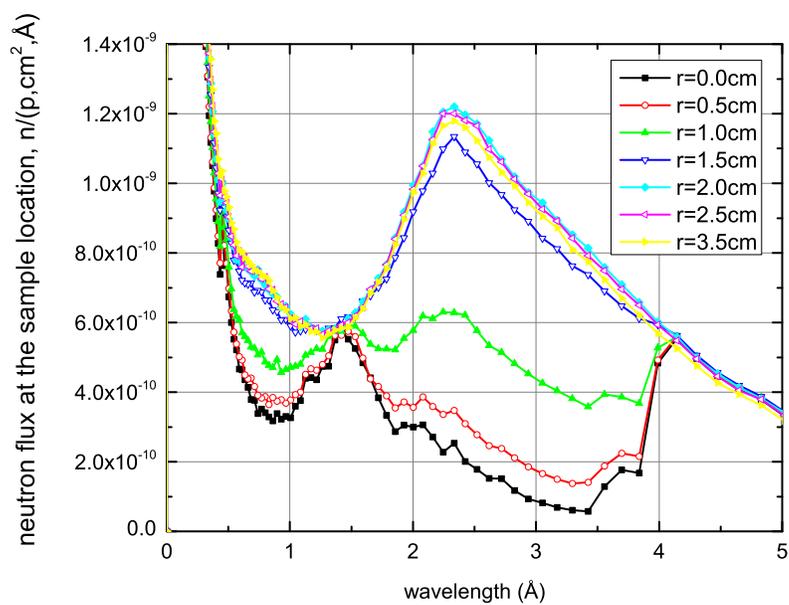}
   \caption{Dependence of the flux at the LQD sample location on the radius of the graphite crystal.}
   \label{fig:radius}
\end{figure}

As shown in section~\ref{sec:Mark-4}, the introduction of a graphite crystal increases the flux between 1.5~\AA~and 4.0~\AA~significantly for the instruments SPEAR and LQD. However, the discontinuity at 4~\AA~will cause problems for the data reduction programs on both instruments. 
It is therefore desirable to eliminate this discontinuity, for example by simply increasing the size of the graphite crystal. This would mean that also the Asterix instrument would receive an increase in the flux below 4~\AA. 
Since the increase would only come in an area that is considered thermal but not in the epi-thermal region, there is a good chance that the instrument responsible for Asterix will accepted these changes. 
Under this assumption the impact of the size of the crystal on the neutron spectrum at the sample location was investigated. 
For this purpose the geometry, as shown in figure~\ref{fig:Mark-4}, was simplified by removing the part of the crystal that points towards the SPEAR instrument. 
The remaining part of the crystal is a cylinder that points towards the LQD instrument.
The radius of this cylinder was varied and results shown in figure~\ref{fig:radius}.
The umbra of the view surface of the moderator has a radius of 1.44~cm for the collimation system used in the LQD instrument model. It can clearly be seen in this figure that for crystal radii below the radius of the umbra the dependence of the flux below 4~\AA~on the crystal radius is very high. Whereas if the radius of the crystal is larger than the umbra radius, the flux at the sample location is almost independent from the radius of the crystal. 
As soon as the size of the crystal is larger than the viewed surface of the instrument, the discontinuity at 4~\AA~disappears. 
The size of the crystal has very little to no impact on the observed epi-thermal flux.

\subsection{Ortho to para hydrogen ratio}

\begin{figure}
   \includegraphics[width=12cm]{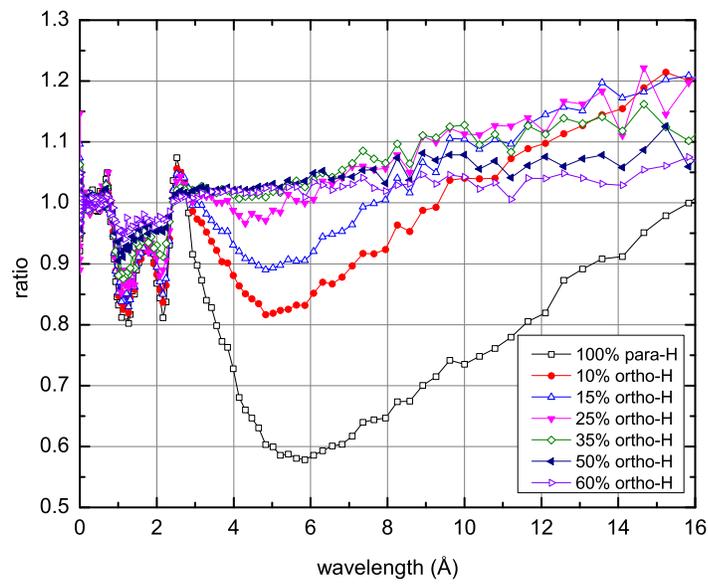}
   \caption{Dependence of the flux at the LQD sample location on the ortho to para hydrogen ratio (normalized to the 75~\% ortho hydrogen spectrum).}
   \label{fig:ortho-para}
\end{figure}

As it has been shown on several occasions in past, the ratio of ortho hydrogen to para hydrogen impacts the performance of a liquid hydrogen moderator significantly~\cite{Ooi2006},~\cite{Muhrer2005}. So far the 
calculations were done using 25~\% ortho hydrogen. 
In order to study the impact of the ortho to para hydrogen ratio independently of the size of the crystal, the size of the graphite was chosen with radius=2~cm. This ensures that the area covered by the crystal is larger than the viewed surface of the LQD instrument. 
The results are displayed in figure~\ref{fig:ortho-para}.
A ortho hydrogen concentration below 25~\% a dependence of the neutron flux on the ortho to para hydrogen ratio is significant. Between 1~\AA~and 3~\AA~the neutron flux decreases with decreasing ortho-hydrogen concentraction, with the flux of pure para-hydrogen being almost 20~\% lower than the one of normal hydrogen (75~\% ortho hydrogen). 

The range of 3~\AA~to 10~\AA~shows the most significant dependence on the ortho to para hydrogen ratio. While in the range of 25~\% to 75~\% ortho-hydrogen the neutron spectrum is almost independent of the ortho to para hydrogen ratio, the magnitude of the spectrum drops significantly with decreasing ortho-hydrogen concentraction below 25~\% ortho-hydrogen. 
In the most extreme case, 6~\AA~and 100~\% para-hydrogen, the flux is about 40~\% lower than the one of normal hydrogen.

In the long wavelength range (10~\AA~to 16~\AA) the flux initially increases with decreasing ortho hydrogen concentration. The maximum flux is reached at about 15~\%-~25~\%. Below 15~\% ortho hydrogen, the flux decreases again. Overall the uncertainty of the integrated flux in this area is up to 20~\%.

\section{Conclusion}
\label{sec:7}

In this paper the theoretical concept of a cold crystal reflector filter was presented. The advantage of this concept over the traditional cold polycrystalline reflector filter lies in the fact that a proper oriented crystal allows for a better transmission of neutron of shorter wavelengths while still preserving the long wavelength neutron flux. It is our intention to benchmark this theoretical concept against a proof of principle experiment.
Hopefully, we will get the opportunity to conduct this experiment during the next run-cycle at the Los Alamos Neutron Science Center. If this proof of principle experiment is successful, we will incoperate this concept in the Mark-4 design of the 1L-target at LANSCE.

\section*{Acknowledgment}

This work was supported by Readiness in Technical Base and Facilities 
(RTBF) which is funded by the Department of Energy's Office of National 
Nuclear Security Administration. It has benefited from the use of the
Manuel Lujan, Jr. Neutron Scattering Center at Los Alamos National 
Laboratory, which is funded by the Department of Energy's Office of Basic
Energy Sciences. Los Alamos National Laboratory is operated by
Los Alamos National Security LLC under DOE Contract
DE-AC52-06NA25396.


\begin{thebibliography}{99}

\bibitem{Carpenter1981} J.M. Carpenter, R. Kleb, T.A. Postal, R.H. Stefiuk, D.F.R. Mildner, Nuclear Instrument and Methods \textbf{189}, 485 (1981).

\bibitem{Kiyanagi1995} Y. Kiyanagi, Y. Ogawa, N. Kosugi, H. Iwasa, M. Furusaka, N. Watanabe, in Proceedings of the 13th Meeting of the International Collaboration on Advanced Neutron Sources, September 11-14, 1995, Villigen, Switzerland.

\bibitem{Pitcher2003} E.J. Pitcher, G.J. Russell, G. Muhrer, J.J. Jarmer, R.K. Corzine, in Proceedings of the 16th Meeting of the International Collaboration on Advanced Neutron Sources, Neuss, Germany (2003).

\bibitem{MCNPX} D.B. Pelowitz et al., MCNPX manual v.2.6.o, LA-CP-07-1473 (2008).

\bibitem{Muhrer2007} G. Muhrer, M.A. Hartl, L.L. Daemen and J. Ryu, Nuclear Instrument and Methods \textbf{578}, 463 (2007).

\bibitem{Muhrer2005} G. Muhrer, E.J. Pitcher and G.J. Russell, Nuclear Instrument and Methods \textbf{536}, 154 (2005).

\bibitem{MockoMark3-pre} M. Mocko and G. Muhrer, in Proceedings of the 18th Meeting of the International Collaboration on Advanced Neutron Sources, Dongguan, China, April 26-29 2007, p. 459.

\bibitem{Lisowski2006} P.W. Lisowski and K.F. Schoenberg, Nuclear Instrument and Methods A \textbf{562}, 910 (2006).

\bibitem{MockoMark3-post} M. Mocko, Ch. T. Kelsey and G. Muhrer, ``Commisioning of the new Lujan TMRS (Mark-III)'', AccApp11, Knoxville, TN, USA, April 3-7 2011.

\bibitem{SPEAR} M. Dubey, P. Wang, M.S. Jablin, M. Mocko and J. Majewski, European Physics Journal Plus \textbf{126}, 110 (2011)

\bibitem{Urban2012} G. Muhrer, Nuclear Instrument and Methods \textbf{664}, 38 (2012).

\bibitem{LQD} P.A.~Seeger, R.P.~Hjelm, M.~Nutter,  Mol. Cryst. Liq. Cryst. \textbf{180A} (1990) 101.

\bibitem{Asterix} lansce.lanl.gov/lujan/instruments/ASTERIX.shtml.

\bibitem{Jarek} J. Majewski (private communication)

\bibitem{njoy1994} R.E. MacFarlane and D.W. Muir, {\it The NJOY Nuclear Data Processing system}, Los Alamos National Laboratory, LA-12740M (Los Alamos 1994).

\bibitem{Gallmeier2012} F.X. Gallmeier (private communication)

\bibitem{Ooi2006} M. Ooi, T. Ino, G. Muhrer, E.J. Pitcher, G.J. Russell, P.D. Ferguson, E.B. Iverson, D. Freeman and Y. Kiyanagi, Nuclear Instrument and Methods \textbf{599}, 699 (2006).


\newpage



















































\end{thebibliography}
\end{document}